\documentclass[twocolumn,superscriptaddress,floatfix,prb]{revtex4-1}

\usepackage{graphicx,epsfig}
\usepackage{SIunits}
\usepackage{miller}
\usepackage{units}
\begin{document}
\title{Electric-field induced switching from fcc to hcp stacking of a single layer of Fe/Ni(111)}
\author{Lukas Gerhard$^{1}$, Moritz Peter$^{2}$, Wulf Wulfhekel$^{1,2}$\\
$^1$Institut f\"ur Nanotechnologie, Karlsruhe Institute of Technology, 76344 Eggenstein-Leopoldshafen, Germany \\
$^2$Physikalisches Institut, Karlsruhe Institute of Technology, 76131 Karlsruhe, Germany \\
}

\begin{abstract}
We present a detailed study of an electric-field induced phase transition of a single layer of Fe on a Ni(111) substrate. Scanning tunneling microscopy at 4 K substrate temperature is used to provide the necessary electric field and to follow the transition from face-centered cubic to hexagonal close-packed stacking with atomic resolution.
\end{abstract}

\maketitle
The concept of electric-field induced switching of magnetic metallic nanosystems is of particular interest in the view of potential applications in data storage devices and led to a tremendous increase of research activities during the last few years \cite{Weisheit2007,Maruyama2009,nnano,Subkow2011,Shiota2012}. In the case of 2 ML Fe islands on Cu(111), it has been shown that a martensitic phase transition between ferromagnetic body-centered cubic(bcc) and antiferromagnetic face-centered cubic (fcc) phases is induced upon application of high electric fields. Herein, the coexistence of two different crystallographic phases and the complex magnetic order are closely interwoven \cite{nnano,fecuheli}. Considering the interatomic distances between nearest neighbors $d_{nn}$ and the resulting lattice mismatch, a Ni(111) surface where $d_{nn}=\unit[248]{pm}$ should be an ideal candidate to promote the coexistence of Fe fcc ($d_{nn}=\unit[253]{pm}$) and bcc ($d_{nn}=\unit[247]{pm}$) films. The first layer of Fe/Ni(111) was considered to nucleate in the hexagonal close-packed (hcp) sites in theoretical calculations and photoelectron diffraction experiments \cite{Longo2007,Gazzadi2002}, while in other experiments, the fcc sites seemed to be favorable \cite{Bradshaw1998}. This fact already hints at a possible coexistence of fcc and hcp phases in the first layers of Fe/Ni(111), which is known for Fe/Ir(111) \citep{Bergmann2007, Heinze2009}. With increasing thickness, a transition from fcc to bcc was reported at thicknesses of 3 to 6 ML \cite{Wu1992, Gazzadi2002}. Altogether this makes the system of Fe/Ni(111) a complex but promising system in the view of possible magnetoelectric coupling (MEC).

We studied the crystallographic structure of Fe films with a local thickness of 1 to 4$\,$ML grown on a Ni(111) single crystal in a low temperature scanning tunneling microscope (STM). STM has been proven to be an ideal tool as it allows the application of extremely high electric fields and at the same time it offers unequaled spacial resolution. Here we fully exploit the potential of STM, directly imaging the transition of a 1$\,$ML Fe film from fcc to hcp stacking with atomic resolution in electric fields of the order of 1$\,$GeV/m.

\begin{figure}
  \centering
 \includegraphics [width = 0.35\textwidth] {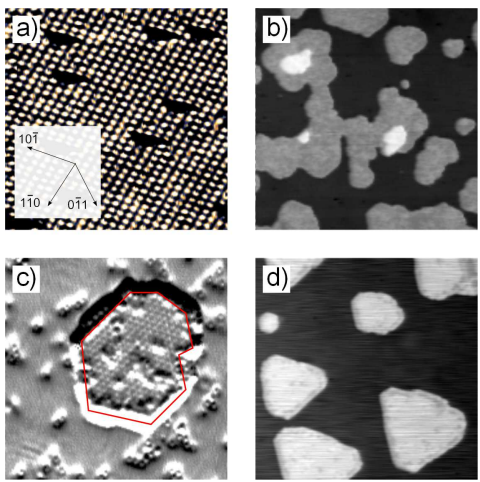}
 \caption{a) Atomically resolved STM topography of the bare Ni(111) single crystal surface, $\unit[7\times7]{nm^2}$. b), d) $\unit[60\times60]{nm^2}$ topography of similar amounts of Fe deposited at about $\unit[30]{\celsius}$: b) Standard preparation leading to a highly contaminated surface and irregular islands with a $2 \times 2$ superstructure (c) induced by adsorbates on an Fe island (enclosed in red) and on the Ni surface. d) Optimized preparation with reduced surface contamination.}
 \label{feni_topo}
 \end{figure}

A Ni single crystal always includes a considerable amount of contaminations and an atomically clean Ni(111) could not be obtained by standard cleaning procedures. After systematical testing, we finally obtained the lowest contamination (about 1 unwanted molecule per 100 Ni atoms, see Fig. \ref{feni_topo}a)) with the following preparation procedure: After repeated cycles of annealing at temperatures $>\unit[700]{\celsius}$ in order to promote segregation of contaminants to the surface and cleaning by Argon sputtering, the last sputtering was short, just enough as to remove residual contamination due to the previous annealing. The final annealing was shortened to about 1 second at $\unit[400]{\celsius}$ to avoid further segregation, to reduce heating of the sample surroundings and thus to minimize cool-down time before deposition of Fe. A cool down is necessary to avoid intermixing of the impinging Fe atoms with the Ni(111) crystal. During this time (about $\unit[1]{\hour}$ at $\unit[1\cdot10^{-10}]{mbar}$), the Ni(111) surface became partly contaminated. Such contamination leads to an irregular growth of the Fe film (see Fig. \ref{feni_topo}b)) and extended clusters with a regular $2\times 2$ superstructure on the bare Ni and particularly on the deposited Fe film (see Fig. \ref{feni_topo}c)). This superstructure can be explained by a hydrogen induced reorganization of the atomic structure as it has been observed by An et al. \cite{An2007} or by carbon contamination which we found by Auger electron spectroscopy (not shown). Cleaner Fe films were obtained by the following measures: 1) For the cool down the Ni(111) sample was placed in the chamber containing the He cryostat which has a lower base pressure of residual gases (e. g. hydrogen). 2) Just before the deposition of Fe, the pressure in the preparation chamber was further improved by sublimation of titanium on the walls cooled with liquid nitrogen. Thereby a base pressure of $\unit[4\cdot10^{-11}]{mbar}$ was achieved during the deposition of Fe. A so-prepared submonolayer Fe film showed a well-ordered morphology with larger terraces and islands oriented along the $\hkl<110>$ directions (see Fig. \ref{feni_topo}d)). 

\begin{figure}
  \centering
 \includegraphics [width = 0.45\textwidth] {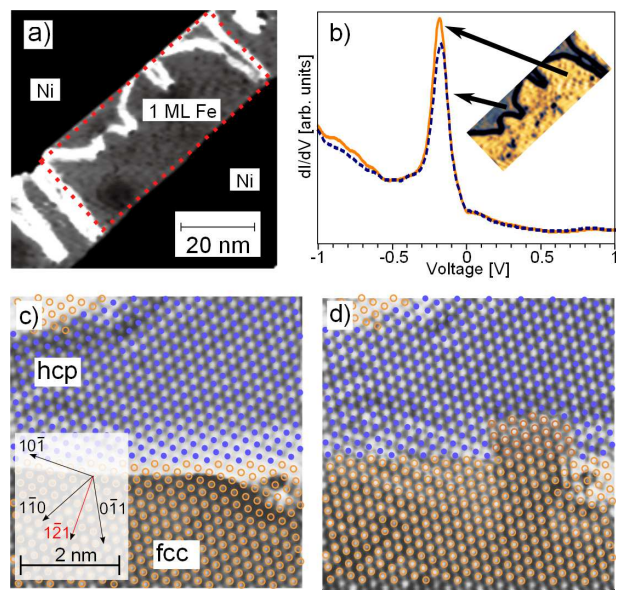}
 \caption{Fe/Ni(111): a) STM topography of a 1 ML Fe film (light gray) at a Ni step edge (black) with domain boundaries appearing in white. b) In a $dI/dV$ map at $\unit[-130]{mV}$ (see inset, position indicated by red rectangle in a)), two different intensities divide the 1 ML Fe terrace into domains. $dI/dV$ spectra outside the domain on the Fe terrace (solid orange) and inside the domain (dashed blue) show different spectra for the two phases. c) Atomic resolution STM showing a coexistence of hcp and fcc lattices with light domain walls. The overlaid hexagonal grid is adjusted to the fcc lattice and colored orange on the fcc area and blue on the hcp area. d) Same area as in c) after an electric field-induced change of the lattice. About 20 atoms change their stacking from hcp to fcc thus shifting the domain boundary to the left by about $\unit[1] {nm}$. 
 } 
 \label{feni_cryst}
 \end{figure}

In topographic measurements on the 1 ML thick Fe films, bright domain boundaries can be seen which enclose irregular domains (see Fig. \ref{feni_cryst}a)). Measurements of the differential conductance $dI/dV$ at energies close to the Fe surface state at $\unit[-200]{mV}$ \cite{Stroscio1995} reveal the white lines in the topography as boundaries between two electronically different domains. As can be seen in a map of the differential conductance ($dI/dV$) taken at $\unit[-130]{mV}$ (see inset in Fig. \ref{feni_cryst}b)) the enclosed domains appear darker. In the $dI/dV$ spectrum, the two phases of 1 ML Fe show different intensities in the negative energy range (see Fig. \ref{feni_cryst}b)). In order to identify the crystallographic structure of the two phases, we performed atomically resolved STM measurements. Fig. \ref{feni_cryst}c) shows an enlarged view of a domain boundary with atomic resolution. The superimposed hexagonal grid is adjusted to the Fe atoms outside the domain (on the lower part of the image) which have a perfectly hexagonal structure and follow the Ni substrate stacking in an fcc order. Within the domain, the fcc grid (blue) does not lie on top of the Fe atoms but exactly in-between on threefold hollow sites. This indicates a hcp stacking of the Fe monolayer with respect to the Ni substrate inside the domain. Hence the observed boundaries are surface dislocations between fcc and hcp domains. About $\unit[10]{\%}$ of the first Fe layer shows hcp stacking with domain boundaries mainly perpendicular to the Ni step edges. This is in agreement with photoelectron diffraction measurements \cite{Theobald1998}. As can be seen in Fig. \ref{feni_cryst}c), the introduction of the hcp domains leads to a reduction of the atomic density in the top layer of about $\unit[1.4]{\%}$. This is equivalent to a reduction of the strain energy of the fcc Fe film. 

To test for MEC, high electric fields were applied by increasing the applied bias voltage. In the 1 ML Fe film, it is possible to induce structural changes by the resulting high electric fields as is shown in Fig. \ref{feni_cryst}: The same field of view is shown before (c) and after (d) an electric-field induced change. It can be seen that a small area of about 20 atoms switched from the hcp domain to the fcc domain thus shifting the domain boundary by about $\unit[1] {nm}$. This transition corresponds to a collective displacement of the Fe atoms by $\unit[143] {pm}$ along the $ \hkl[1-21]_{fcc}$ direction from the hcp threefold hollow sites to the fcc positions. The new domain boundary in this direction is confined to only one interatomic distance while it extends over 3 to 5 interatomic distances in the pristine configuration. 

\begin{figure}
  \centering
\includegraphics[width = 0.45\textwidth]{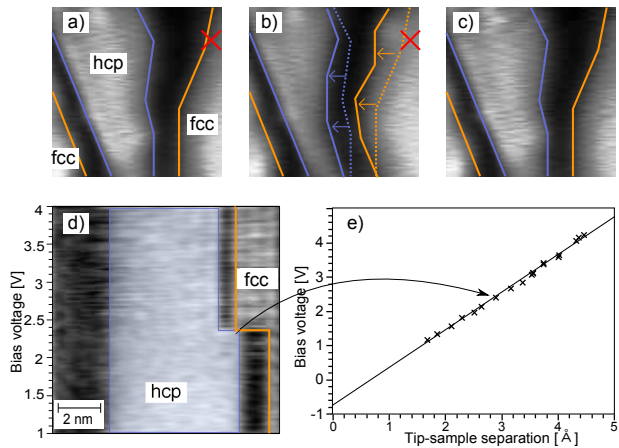}
 \caption{Reproducible switching of the crystallographic structure of 1 ML Fe/Ni(111) imaged on $\unit[6.25\times6.25]{nm^2}$ ($dI/dV$ map at $\unit[-190]{mV}$ ). a) The initial hcp domain is indicated by a blue line, the fcc domain by a orange line, domain boundaries appear dark. An electric field pulse is applied on the position marked with the red cross after the acquisition of the image. The new configuration with a smaller hcp domain is imaged in b) with the initial position of the domains indicated by dashed lines. A second pulse brings the system back to the initial state (c)). Critical parameters for the transition: d) The same line across a hcp domain (blue frame) is repeatedly scanned while the voltage is ramped.  Above a certain critical voltage (here $\unit[2.3]{V}$), the structure partially changes from hcp to fcc and the domain boundary (dark) is shifted to the right. Repetition of this measurement at different tip-sample separations gives the critical electric field shown in e). The measurement described in d) corresponds to one data point (black cross). The linear relation between the critical voltage and the tip-sample separation corresponds to a critical electric field of about $\unitfrac[10]{GV}{m}$.} 
 \label{feni_mec_repro}
\end{figure}

In a further experiment the bistability of the switching was confirmed. An hcp domain was imaged at a moderate electric field of $\unitfrac[0.6]{GV}{m}$ (see Fig. \ref{feni_mec_repro}a)). The domain boundaries appear dark and the hcp and the fcc domains are marked with blue and orange. A high electric field pulse of $\unitfrac[7.9]{GV}{m}$ and $\unit[0.1]{s}$ was applied at the position marked with the red cross as to locally change the structure. A second STM image at $\unitfrac[0.6]{GV}{m}$ shows the manipulated configuration with a reduced hcp domain (see Fig.\ref{feni_mec_repro}b)). After a second electric field pulse, the original situation was restored (see Fig. \ref{feni_mec_repro}c)). 
 
In order to prove that the observed switching is induced by the applied electric field, we performed a systematic study varying the applied voltage and the distance between tip and sample. For each distance, which we estimated from the tunneling current and bias voltage, the same line across the hcp domain was scanned repeatedly while the bias voltage was increased. This allows to measure the critical value of the voltage at which the domain boundary is displaced (see Fig. \ref{feni_mec_repro}d)). We repeated this experiment for a set of distances and the measured critical values can be nicely fitted by a linear distance dependence of the voltage as it is expected for a critical electric field. This excludes other mechanisms than the electric field because they would result in distinctly different critical-voltage-distance relations \cite{nnano}. 
The negative offset of the voltage for zero distance can be explained by the difference of the work functions of the tungsten tip and the Fe sample surface which induces a potential difference even without an externally applied bias.
The slope of the linear fit gives the critical value of the electric field of about $\unitfrac[10]{GV}{m}$ (see Fig.\ref{feni_mec_repro}e)). This value is about one order of magnitude larger than the electric fields needed for MEC in 2 ML Fe/Cu(111) \cite{nnano,PRL2013}. These high electric fields require a very stable configuration of the tip apex. 

\begin{figure}
\centering
\includegraphics[width = 0.45\textwidth]{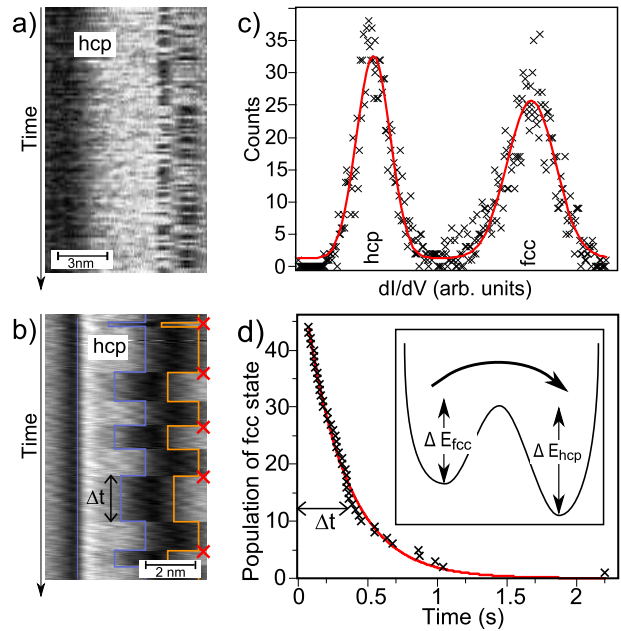}
 \caption{a) Switching the domain boundary left and right when scanning across an hcp domain at a constant field of $\unitfrac[7.6]{GV}{m}$ ($dI/dV$ at $3.7\,V$). b) Switching left by field pulses of $\unitfrac[1.4]{GV}{m}$ and right by scanning at $\unitfrac[1.0]{GV}{m}$ across an hcp domain (blue) ($dI/dV$ at $\unit[-240]{mV}$). c) In a histogram of the differential conductance, only two stable states are found. d) Exponential decay of the fcc state with a lifetime of $\unit[0.3]{s}$.} 
\label{feni_dynamic}
\end{figure}

Similar to the experiments on Fe/Cu(111), electric fields slightly lower than the critical values lead to switching after a certain residence time \cite{PRL2013}. This can be seen in Fig. \ref{feni_dynamic}a) which shows switching left and right of the domain boundary during scanning at a constant electric field of $\unitfrac[7.6]{GV}{m}$. However, this implies a high bias voltage (in this case $\unit[3]{V}$) which leads to a reduced contrast and a rather low signal-to-noise ratio. Therefore, the distribution of the residence times $\Delta t$ is studied as follows (see Fig. \ref{feni_dynamic}b)): the displacement to the left is induced by electric field pulses of $\unitfrac[1.4]{GV}{m}$ during $\unit[0.1]{s}$ (at the positions indicated in the Fig.), while the switching back to the right happens after scanning at $\unitfrac[1.0]{GV}{m}$ for some time. A displacement to the right has never been observed during scanning at this value. The corresponding histogram of the measured $dI/dV$ intensity at the initial position of the 
domain boundary shows only two different peaks, corresponding to two stable positions of the domain boundary. The plot of the residence times of the hcp state is depicted in Fig. \ref{feni_dynamic}d) and shows an exponential decay with a lifetime $\tau_{fcc}$ of about $\unit[0.3]{s}$. In analogy to our experiments on Fe/Cu(111) reported in \citep{PRL2013}, this can be explained by a thermally activated transition between two local energy minima (see inset in Fig. \ref{feni_dynamic}d).
In contrast to our findings for Fe/Cu(111)\cite{fecustat}, a clear relation between the polarity of the electric field and the resulting lattice state could not be found. Switching in both directions was possible with pulses of any polarity with approximately the same probability.

\begin{figure}
  \centering
 \includegraphics [width = 0.45\textwidth] {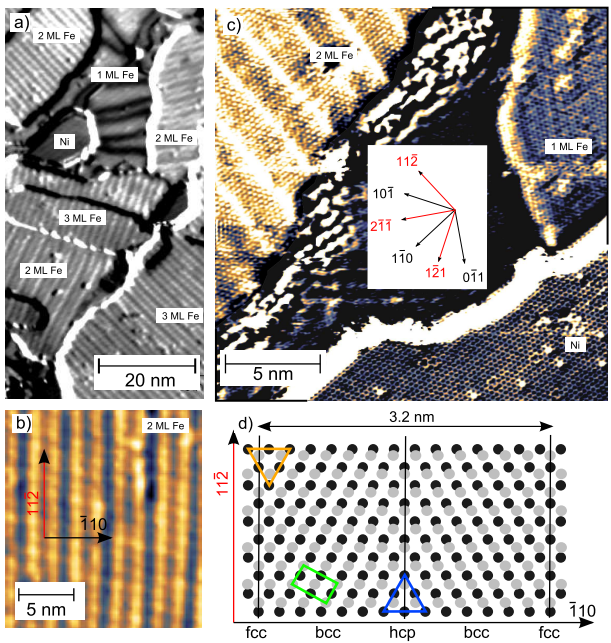}
 \caption{a) A  $dI/dV$ map at $\unit[-240]{mV}$ of Fe/Ni(111) films thicker than 1 ML shows a stripe pattern perpendicular to the closed-packed $\hkl<110>$ directions. The lattice directions given in c) also hold for a). b) Three different intensities of the stripes are found. c) STM image on the bare Ni surface, $\unit[1]{ML}$ Fe and $\unit[2]{ML}$ Fe with atomic resolution. d) Atomic model of 2 ML Fe/Ni(111) for the observed stripe pattern. The relative atomic distances of the top layer (gray circles) are taken from the measurement in c). The expansion of the top layer by $\unit[7.5]{\%}$ in the $\hkl[1-10]$ direction leads to a periodicity of 13 interatomic distances ($\unit[3.2] {nm}$). The bcc lattice (green rectangle) is characterized by a twofold bridge-stacking while the two possible threefold hollow sites correspond to fcc (orange triangle) and hcp domains (blue triangle).} 
 \label{feni_2fe}
 \end{figure}

Deposition of higher amounts of Fe resulted in films locally 2 and more ML thick, showing a superstructure with periodic stripes aligned perpendicular to the three closed packed $\hkl<110>$ directions, in full agreement with the paper from An et al. \cite{An2009}. They observed bright and dark stripes in STM and two satellite spots in the corresponding LEED pattern. This was interpreted as a bcc $\hkl(110)$ lattice in the Nishiyama-Wassermann orientation that is expanded in the $\hkl[1-10]_{bcc}$ direction by $\unit[6.5]{\%}$. In our STM measurements, we observed the same stripe pattern (see Fig. \ref{feni_2fe}a)). However, in our $dI/dV$ maps, we identify three different intensities of the stripes and thus find a doubled periodicity (see Fig. \ref{feni_2fe}b)), which is in contradiction to the model proposed by An et al. An atomically resolved STM image of both the Ni surface and the $\unit[2]{ML}$ Fe film (see Fig. \ref{feni_2fe}c)) allows us to unambiguously determine the atomic structure: 
The $\unit[2]{ML}$ Fe film is expanded by $\unit[7.5]{\%}$ in the $\hkl[-110]_{fcc}$ direction and by $\unit[1.9]{\%}$ in the $ \hkl[11-2]_{fcc}$ direction with respect to the Ni lattice. Fig. \ref{feni_2fe}d) proposes an atomic model that explains the experimentally observed features by a repetition of fcc, bcc, hcp, bcc-like stripes with a periodicity of $\unit[3.2] {nm}$. The fcc and hcp areas appear dark, the bcc-like areas appear bright (compare to Fig. \ref{feni_2fe}b)). These structures of the Fe films thicker than 1 ML could not be influenced by the applied electric field.
 
In summary, we have identified the complex crystallographic phase diagram of Fe/Ni(111) by atomic resolution STM. While Fe films of 2 and more layers show periodic stripes of fcc, bcc and hcp phases, the single monolayer shows domains of fcc and hcp stacking. By application of high electric fields, we have been able to induce a reproducible transition from fcc the stacking to hcp. The underlying mechanism of this transition is similar to that previously observed in 2 ML Fe/Cu(111) and thus seems to be a more universal concept for metallic surfaces that is not limited to a particular system.

We acknowledge funding by the DFG under gran WU 349/8-1.

\bibliographystyle{aapmrev4-1}
\bibliography{feni}

\begin{thebibliography}{18}%
\makeatletter
\providecommand \@ifxundefined [1]{%
 \@ifx{#1\undefined}
}%
\providecommand \@ifnum [1]{%
 \ifnum #1\expandafter \@firstoftwo
 \else \expandafter \@secondoftwo
 \fi
}%
\providecommand \@ifx [1]{%
 \ifx #1\expandafter \@firstoftwo
 \else \expandafter \@secondoftwo
 \fi
}%
\providecommand \natexlab [1]{#1}%
\providecommand \enquote  [1]{``#1''}%
\providecommand \bibnamefont  [1]{#1}%
\providecommand \bibfnamefont [1]{#1}%
\providecommand \citenamefont [1]{#1}%
\providecommand \href@noop [0]{\@secondoftwo}%
\providecommand \href [0]{\begingroup \@sanitize@url \@href}%
\providecommand \@href[1]{\@@startlink{#1}\@@href}%
\providecommand \@@href[1]{\endgroup#1\@@endlink}%
\providecommand \@sanitize@url [0]{\catcode `\\12\catcode `\$12\catcode
  `\&12\catcode `\#12\catcode `\^12\catcode `\_12\catcode `\%12\relax}%
\providecommand \@@startlink[1]{}%
\providecommand \@@endlink[0]{}%
\providecommand \url  [0]{\begingroup\@sanitize@url \@url }%
\providecommand \@url [1]{\endgroup\@href {#1}{\urlprefix }}%
\providecommand \urlprefix  [0]{URL }%
\providecommand \Eprint [0]{\href }%
\providecommand \doibase [0]{http://dx.doi.org/}%
\providecommand \selectlanguage [0]{\@gobble}%
\providecommand \bibinfo  [0]{\@secondoftwo}%
\providecommand \bibfield  [0]{\@secondoftwo}%
\providecommand \translation [1]{[#1]}%
\providecommand \BibitemOpen [0]{}%
\providecommand \bibitemStop [0]{}%
\providecommand \bibitemNoStop [0]{.\EOS\space}%
\providecommand \EOS [0]{\spacefactor3000\relax}%
\providecommand \BibitemShut  [1]{\csname bibitem#1\endcsname}%
\let\auto@bib@innerbib\@empty
\bibitem [{\citenamefont {Weisheit}\ \emph {et~al.}(2007)\citenamefont
  {Weisheit}, \citenamefont {F\"{a}hler}, \citenamefont {Marty}, \citenamefont
  {Souche}, \citenamefont {Poinsignon},\ and\ \citenamefont
  {Givord}}]{Weisheit2007}%
  \BibitemOpen
  \bibfield  {author} {\bibinfo {author} {\bibfnamefont {M.}~\bibnamefont
  {Weisheit}}, \bibinfo {author} {\bibfnamefont {S.}~\bibnamefont
  {F\"{a}hler}}, \bibinfo {author} {\bibfnamefont {A.}~\bibnamefont {Marty}},
  \bibinfo {author} {\bibfnamefont {Y.}~\bibnamefont {Souche}}, \bibinfo
  {author} {\bibfnamefont {C.}~\bibnamefont {Poinsignon}}, \ and\ \bibinfo
  {author} {\bibfnamefont {D.}~\bibnamefont {Givord}},\ }\href {\doibase
  10.1126/science.1136629} {\bibfield  {journal} {\bibinfo  {journal}
  {Science}\ }\textbf {\bibinfo {volume} {315}},\ \bibinfo {pages} {349}
  (\bibinfo {year} {2007})}\BibitemShut {NoStop}%
\bibitem [{\citenamefont {Maruyama}\ \emph {et~al.}(2009)\citenamefont
  {Maruyama}, \citenamefont {Shiota}, \citenamefont {Nozaki}, \citenamefont
  {Ohta}, \citenamefont {Toda}, \citenamefont {Mizuguchi}, \citenamefont
  {Tulapurkar}, \citenamefont {Shinjo}, \citenamefont {Shiraishi},
  \citenamefont {Mizukami}, \citenamefont {Ando},\ and\ \citenamefont
  {Suzuki}}]{Maruyama2009}%
  \BibitemOpen
  \bibfield  {author} {\bibinfo {author} {\bibfnamefont {T.}~\bibnamefont
  {Maruyama}}, \bibinfo {author} {\bibfnamefont {Y.}~\bibnamefont {Shiota}},
  \bibinfo {author} {\bibfnamefont {T.}~\bibnamefont {Nozaki}}, \bibinfo
  {author} {\bibfnamefont {K.}~\bibnamefont {Ohta}}, \bibinfo {author}
  {\bibfnamefont {N.}~\bibnamefont {Toda}}, \bibinfo {author} {\bibfnamefont
  {M.}~\bibnamefont {Mizuguchi}}, \bibinfo {author} {\bibfnamefont {A.~A.}\
  \bibnamefont {Tulapurkar}}, \bibinfo {author} {\bibfnamefont
  {T.}~\bibnamefont {Shinjo}}, \bibinfo {author} {\bibfnamefont
  {M.}~\bibnamefont {Shiraishi}}, \bibinfo {author} {\bibfnamefont
  {S.}~\bibnamefont {Mizukami}}, \bibinfo {author} {\bibfnamefont
  {Y.}~\bibnamefont {Ando}}, \ and\ \bibinfo {author} {\bibfnamefont
  {Y.}~\bibnamefont {Suzuki}},\ }\href {\doibase 10.1038/NNANO.2008.406}
  {\bibfield  {journal} {\bibinfo  {journal} {Nat. Nanotechnol.}\ }\textbf
  {\bibinfo {volume} {4}},\ \bibinfo {pages} {158} (\bibinfo {year}
  {2009})}\BibitemShut {NoStop}%
\bibitem [{\citenamefont {Gerhard}\ \emph {et~al.}(2010)\citenamefont
  {Gerhard}, \citenamefont {Yamada}, \citenamefont {Balashov}, \citenamefont
  {Tak\'acs}, \citenamefont {Wesselink}, \citenamefont {D\"ane}, \citenamefont
  {Fechner}, \citenamefont {Ostanin}, \citenamefont {Ernst}, \citenamefont
  {Mertig},\ and\ \citenamefont {Wulfhekel}}]{nnano}%
  \BibitemOpen
  \bibfield  {author} {\bibinfo {author} {\bibfnamefont {L.}~\bibnamefont
  {Gerhard}}, \bibinfo {author} {\bibfnamefont {T.~K.}\ \bibnamefont {Yamada}},
  \bibinfo {author} {\bibfnamefont {T.}~\bibnamefont {Balashov}}, \bibinfo
  {author} {\bibfnamefont {A.~F.}\ \bibnamefont {Tak\'acs}}, \bibinfo {author}
  {\bibfnamefont {R.~J.~H.}\ \bibnamefont {Wesselink}}, \bibinfo {author}
  {\bibfnamefont {M.}~\bibnamefont {D\"ane}}, \bibinfo {author} {\bibfnamefont
  {M.}~\bibnamefont {Fechner}}, \bibinfo {author} {\bibfnamefont
  {S.}~\bibnamefont {Ostanin}}, \bibinfo {author} {\bibfnamefont
  {A.}~\bibnamefont {Ernst}}, \bibinfo {author} {\bibfnamefont
  {I.}~\bibnamefont {Mertig}}, \ and\ \bibinfo {author} {\bibfnamefont
  {W.}~\bibnamefont {Wulfhekel}},\ }\href {\doibase 10.1038/nnano.2010.214}
  {\bibfield  {journal} {\bibinfo  {journal} {Nat. Nanotechnol.}\ }\textbf
  {\bibinfo {volume} {5}},\ \bibinfo {pages} {792} (\bibinfo {year}
  {2010})}\BibitemShut {NoStop}%
\bibitem [{\citenamefont {Subkow}\ and\ \citenamefont
  {F\"ahnle}(2011)}]{Subkow2011}%
  \BibitemOpen
  \bibfield  {author} {\bibinfo {author} {\bibfnamefont {S.}~\bibnamefont
  {Subkow}}\ and\ \bibinfo {author} {\bibfnamefont {M.}~\bibnamefont
  {F\"ahnle}},\ }\href {\doibase 10.1103/PhysRevB.84.054443} {\bibfield
  {journal} {\bibinfo  {journal} {Phys. Rev. B}\ }\textbf {\bibinfo {volume}
  {84}},\ \bibinfo {pages} {054443} (\bibinfo {year} {2011})}\BibitemShut
  {NoStop}%
\bibitem [{\citenamefont {Shiota}\ \emph {et~al.}(2012)\citenamefont {Shiota},
  \citenamefont {Nozaki}, \citenamefont {Bonell}, \citenamefont {Murakami},
  \citenamefont {Shinjo},\ and\ \citenamefont {Suzuki}}]{Shiota2012}%
  \BibitemOpen
  \bibfield  {author} {\bibinfo {author} {\bibfnamefont {Y.}~\bibnamefont
  {Shiota}}, \bibinfo {author} {\bibfnamefont {T.}~\bibnamefont {Nozaki}},
  \bibinfo {author} {\bibfnamefont {F.}~\bibnamefont {Bonell}}, \bibinfo
  {author} {\bibfnamefont {S.}~\bibnamefont {Murakami}}, \bibinfo {author}
  {\bibfnamefont {T.}~\bibnamefont {Shinjo}}, \ and\ \bibinfo {author}
  {\bibfnamefont {Y.}~\bibnamefont {Suzuki}},\ }\href {\doibase
  10.1038/nmat3172} {\bibfield  {journal} {\bibinfo  {journal} {Nat. Mater.}\
  }\textbf {\bibinfo {volume} {11}},\ \bibinfo {pages} {39} (\bibinfo {year}
  {2012})}\BibitemShut {NoStop}%
\bibitem [{\citenamefont {Phark}\ \emph {et~al.}(2014)\citenamefont {Phark},
  \citenamefont {Fischer}, \citenamefont {Corbetta}, \citenamefont {Sander},
  \citenamefont {Nakamura},\ and\ \citenamefont {J.}}]{fecuheli}%
  \BibitemOpen
  \bibfield  {author} {\bibinfo {author} {\bibfnamefont {S.~H.}\ \bibnamefont
  {Phark}}, \bibinfo {author} {\bibfnamefont {J.~A.}\ \bibnamefont {Fischer}},
  \bibinfo {author} {\bibfnamefont {M.}~\bibnamefont {Corbetta}}, \bibinfo
  {author} {\bibfnamefont {D.}~\bibnamefont {Sander}}, \bibinfo {author}
  {\bibfnamefont {K.}~\bibnamefont {Nakamura}}, \ and\ \bibinfo {author}
  {\bibfnamefont {K.}~\bibnamefont {J.}},\ }\href@noop {} {\bibfield  {journal}
  {\bibinfo  {journal} {Nat. Comms.}\ }\textbf {\bibinfo {volume} {5}},\
  \bibinfo {pages} {5183} (\bibinfo {year} {2014})}\BibitemShut {NoStop}%
\bibitem [{\citenamefont {Longo}\ \emph {et~al.}(2007)\citenamefont {Longo},
  \citenamefont {Martinez}, \citenamefont {Dieguez}, \citenamefont {Vega},\
  and\ \citenamefont {Gallego}}]{Longo2007}%
  \BibitemOpen
  \bibfield  {author} {\bibinfo {author} {\bibfnamefont {R.~C.}\ \bibnamefont
  {Longo}}, \bibinfo {author} {\bibfnamefont {E.}~\bibnamefont {Martinez}},
  \bibinfo {author} {\bibfnamefont {O.}~\bibnamefont {Dieguez}}, \bibinfo
  {author} {\bibfnamefont {A.}~\bibnamefont {Vega}}, \ and\ \bibinfo {author}
  {\bibfnamefont {L.~J.}\ \bibnamefont {Gallego}},\ }\href
  {http://stacks.iop.org/0957-4484/18/i=5/a=055701} {\bibfield  {journal}
  {\bibinfo  {journal} {Nanotechnology}\ }\textbf {\bibinfo {volume} {18}},\
  \bibinfo {pages} {055701} (\bibinfo {year} {2007})}\BibitemShut {NoStop}%
\bibitem [{\citenamefont {Gazzadi}\ \emph {et~al.}(2002)\citenamefont
  {Gazzadi}, \citenamefont {Bruno}, \citenamefont {Capelli}, \citenamefont
  {Pasquali},\ and\ \citenamefont {Nannarone}}]{Gazzadi2002}%
  \BibitemOpen
  \bibfield  {author} {\bibinfo {author} {\bibfnamefont {G.~C.}\ \bibnamefont
  {Gazzadi}}, \bibinfo {author} {\bibfnamefont {F.}~\bibnamefont {Bruno}},
  \bibinfo {author} {\bibfnamefont {R.}~\bibnamefont {Capelli}}, \bibinfo
  {author} {\bibfnamefont {L.}~\bibnamefont {Pasquali}}, \ and\ \bibinfo
  {author} {\bibfnamefont {S.}~\bibnamefont {Nannarone}},\ }\href {\doibase
  10.1103/PhysRevB.65.205417} {\bibfield  {journal} {\bibinfo  {journal} {Phys.
  Rev. B}\ }\textbf {\bibinfo {volume} {65}},\ \bibinfo {pages} {205417}
  (\bibinfo {year} {2002})}\BibitemShut {NoStop}%
\bibitem [{\citenamefont {Theobald}\ \emph
  {et~al.}(1998{\natexlab{a}})\citenamefont {Theobald}, \citenamefont
  {Fernandez}, \citenamefont {Schaff}, \citenamefont {Hofmann}, \citenamefont
  {Schindler}, \citenamefont {Fritzsche}, \citenamefont {Bradshaw},\ and\
  \citenamefont {Woodruff}}]{Bradshaw1998}%
  \BibitemOpen
  \bibfield  {author} {\bibinfo {author} {\bibfnamefont {A.}~\bibnamefont
  {Theobald}}, \bibinfo {author} {\bibfnamefont {V.}~\bibnamefont {Fernandez}},
  \bibinfo {author} {\bibfnamefont {O.}~\bibnamefont {Schaff}}, \bibinfo
  {author} {\bibfnamefont {P.}~\bibnamefont {Hofmann}}, \bibinfo {author}
  {\bibfnamefont {K.-M.}\ \bibnamefont {Schindler}}, \bibinfo {author}
  {\bibfnamefont {V.}~\bibnamefont {Fritzsche}}, \bibinfo {author}
  {\bibfnamefont {A.~M.}\ \bibnamefont {Bradshaw}}, \ and\ \bibinfo {author}
  {\bibfnamefont {D.~P.}\ \bibnamefont {Woodruff}},\ }\href {\doibase
  10.1103/PhysRevB.58.6768} {\bibfield  {journal} {\bibinfo  {journal} {Phys.
  Rev. B}\ }\textbf {\bibinfo {volume} {58}},\ \bibinfo {pages} {6768}
  (\bibinfo {year} {1998}{\natexlab{a}})}\BibitemShut {NoStop}%
\bibitem [{\citenamefont {von Bergmann}\ \emph {et~al.}(2007)\citenamefont {von
  Bergmann}, \citenamefont {Heinze}, \citenamefont {Bode}, \citenamefont
  {Bihlmayer}, \citenamefont {Blügel},\ and\ \citenamefont
  {Wiesendanger}}]{Bergmann2007}%
  \BibitemOpen
  \bibfield  {author} {\bibinfo {author} {\bibfnamefont {K.}~\bibnamefont {von
  Bergmann}}, \bibinfo {author} {\bibfnamefont {S.}~\bibnamefont {Heinze}},
  \bibinfo {author} {\bibfnamefont {M.}~\bibnamefont {Bode}}, \bibinfo {author}
  {\bibfnamefont {G.}~\bibnamefont {Bihlmayer}}, \bibinfo {author}
  {\bibfnamefont {S.}~\bibnamefont {Blügel}}, \ and\ \bibinfo {author}
  {\bibfnamefont {R.}~\bibnamefont {Wiesendanger}},\ }\href
  {http://stacks.iop.org/1367-2630/9/i=10/a=396} {\bibfield  {journal}
  {\bibinfo  {journal} {New Journal of Physics}\ }\textbf {\bibinfo {volume}
  {9}},\ \bibinfo {pages} {396} (\bibinfo {year} {2007})}\BibitemShut {NoStop}%
\bibitem [{\citenamefont {Hardrat}\ \emph {et~al.}(2009)\citenamefont
  {Hardrat}, \citenamefont {Al-Zubi}, \citenamefont {Ferriani}, \citenamefont
  {Bl\"ugel}, \citenamefont {Bihlmayer},\ and\ \citenamefont
  {Heinze}}]{Heinze2009}%
  \BibitemOpen
  \bibfield  {author} {\bibinfo {author} {\bibfnamefont {B.}~\bibnamefont
  {Hardrat}}, \bibinfo {author} {\bibfnamefont {A.}~\bibnamefont {Al-Zubi}},
  \bibinfo {author} {\bibfnamefont {P.}~\bibnamefont {Ferriani}}, \bibinfo
  {author} {\bibfnamefont {S.}~\bibnamefont {Bl\"ugel}}, \bibinfo {author}
  {\bibfnamefont {G.}~\bibnamefont {Bihlmayer}}, \ and\ \bibinfo {author}
  {\bibfnamefont {S.}~\bibnamefont {Heinze}},\ }\href {\doibase
  10.1103/PhysRevB.79.094411} {\bibfield  {journal} {\bibinfo  {journal} {Phys.
  Rev. B}\ }\textbf {\bibinfo {volume} {79}},\ \bibinfo {pages} {094411}
  (\bibinfo {year} {2009})}\BibitemShut {NoStop}%
\bibitem [{\citenamefont {Wu}\ and\ \citenamefont {Freeman}(1992)}]{Wu1992}%
  \BibitemOpen
  \bibfield  {author} {\bibinfo {author} {\bibfnamefont {R.}~\bibnamefont
  {Wu}}\ and\ \bibinfo {author} {\bibfnamefont {A.~J.}\ \bibnamefont
  {Freeman}},\ }\href {\doibase 10.1103/PhysRevB.45.7205} {\bibfield  {journal}
  {\bibinfo  {journal} {Phys. Rev. B}\ }\textbf {\bibinfo {volume} {45}},\
  \bibinfo {pages} {7205} (\bibinfo {year} {1992})}\BibitemShut {NoStop}%
\bibitem [{\citenamefont {An}\ \emph {et~al.}(2007)\citenamefont {An},
  \citenamefont {Zhang}, \citenamefont {Fukuyama},\ and\ \citenamefont
  {Yokogawa}}]{An2007}%
  \BibitemOpen
  \bibfield  {author} {\bibinfo {author} {\bibfnamefont {B.}~\bibnamefont
  {An}}, \bibinfo {author} {\bibfnamefont {L.}~\bibnamefont {Zhang}}, \bibinfo
  {author} {\bibfnamefont {S.}~\bibnamefont {Fukuyama}}, \ and\ \bibinfo
  {author} {\bibfnamefont {K.}~\bibnamefont {Yokogawa}},\ }\href {\doibase
  10.1143/JJAP.46.5586} {\bibfield  {journal} {\bibinfo  {journal} {Japanese
  Journal of Applied Physics}\ }\textbf {\bibinfo {volume} {46}},\ \bibinfo
  {pages} {5586} (\bibinfo {year} {2007})}\BibitemShut {NoStop}%
\bibitem [{\citenamefont {Stroscio}\ \emph {et~al.}(1995)\citenamefont
  {Stroscio}, \citenamefont {Pierce}, \citenamefont {Davies}, \citenamefont
  {Celotta},\ and\ \citenamefont {Weinert}}]{Stroscio1995}%
  \BibitemOpen
  \bibfield  {author} {\bibinfo {author} {\bibfnamefont {J.~A.}\ \bibnamefont
  {Stroscio}}, \bibinfo {author} {\bibfnamefont {D.~T.}\ \bibnamefont
  {Pierce}}, \bibinfo {author} {\bibfnamefont {A.}~\bibnamefont {Davies}},
  \bibinfo {author} {\bibfnamefont {R.~J.}\ \bibnamefont {Celotta}}, \ and\
  \bibinfo {author} {\bibfnamefont {M.}~\bibnamefont {Weinert}},\ }\href
  {\doibase 10.1103/PhysRevLett.75.2960} {\bibfield  {journal} {\bibinfo
  {journal} {Phys. Rev. Lett.}\ }\textbf {\bibinfo {volume} {75}},\ \bibinfo
  {pages} {2960} (\bibinfo {year} {1995})}\BibitemShut {NoStop}%
\bibitem [{\citenamefont {Theobald}\ \emph
  {et~al.}(1998{\natexlab{b}})\citenamefont {Theobald}, \citenamefont
  {Fernandez}, \citenamefont {Schaff}, \citenamefont {Hofmann}, \citenamefont
  {Schindler}, \citenamefont {Fritzsche}, \citenamefont {Bradshaw},\ and\
  \citenamefont {Woodruff}}]{Theobald1998}%
  \BibitemOpen
  \bibfield  {author} {\bibinfo {author} {\bibfnamefont {A.}~\bibnamefont
  {Theobald}}, \bibinfo {author} {\bibfnamefont {V.}~\bibnamefont {Fernandez}},
  \bibinfo {author} {\bibfnamefont {O.}~\bibnamefont {Schaff}}, \bibinfo
  {author} {\bibfnamefont {P.}~\bibnamefont {Hofmann}}, \bibinfo {author}
  {\bibfnamefont {K.-M.}\ \bibnamefont {Schindler}}, \bibinfo {author}
  {\bibfnamefont {V.}~\bibnamefont {Fritzsche}}, \bibinfo {author}
  {\bibfnamefont {A.~M.}\ \bibnamefont {Bradshaw}}, \ and\ \bibinfo {author}
  {\bibfnamefont {D.~P.}\ \bibnamefont {Woodruff}},\ }\href {\doibase
  10.1103/PhysRevB.58.6768} {\bibfield  {journal} {\bibinfo  {journal} {Phys.
  Rev. B}\ }\textbf {\bibinfo {volume} {58}},\ \bibinfo {pages} {6768}
  (\bibinfo {year} {1998}{\natexlab{b}})}\BibitemShut {NoStop}%
\bibitem [{\citenamefont {Gerhard}\ \emph {et~al.}(2013)\citenamefont
  {Gerhard}, \citenamefont {Wesselink}, \citenamefont {Ostanin}, \citenamefont
  {Ernst},\ and\ \citenamefont {Wulfhekel}}]{PRL2013}%
  \BibitemOpen
  \bibfield  {author} {\bibinfo {author} {\bibfnamefont {L.}~\bibnamefont
  {Gerhard}}, \bibinfo {author} {\bibfnamefont {R.~J.~H.}\ \bibnamefont
  {Wesselink}}, \bibinfo {author} {\bibfnamefont {S.}~\bibnamefont {Ostanin}},
  \bibinfo {author} {\bibfnamefont {A.}~\bibnamefont {Ernst}}, \ and\ \bibinfo
  {author} {\bibfnamefont {W.}~\bibnamefont {Wulfhekel}},\ }\href {\doibase
  10.1103/PhysRevLett.111.167601} {\bibfield  {journal} {\bibinfo  {journal}
  {Phys. Rev. Lett.}\ }\textbf {\bibinfo {volume} {111}},\ \bibinfo {pages}
  {167601} (\bibinfo {year} {2013})}\BibitemShut {NoStop}%
\bibitem [{\citenamefont {Gerhard}\ \emph {et~al.}(2011)\citenamefont
  {Gerhard}, \citenamefont {Yamada}, \citenamefont {Balashov}, \citenamefont
  {Takacs}, \citenamefont {Wesselink}, \citenamefont {D\"ane}, \citenamefont
  {Fechner}, \citenamefont {Ostanin}, \citenamefont {Ernst}, \citenamefont
  {Mertig},\ and\ \citenamefont {Wulfhekel}}]{fecustat}%
  \BibitemOpen
  \bibfield  {author} {\bibinfo {author} {\bibfnamefont {L.}~\bibnamefont
  {Gerhard}}, \bibinfo {author} {\bibfnamefont {T.}~\bibnamefont {Yamada}},
  \bibinfo {author} {\bibfnamefont {T.}~\bibnamefont {Balashov}}, \bibinfo
  {author} {\bibfnamefont {A.}~\bibnamefont {Takacs}}, \bibinfo {author}
  {\bibfnamefont {R.}~\bibnamefont {Wesselink}}, \bibinfo {author}
  {\bibfnamefont {M.}~\bibnamefont {D\"ane}}, \bibinfo {author} {\bibfnamefont
  {M.}~\bibnamefont {Fechner}}, \bibinfo {author} {\bibfnamefont
  {S.}~\bibnamefont {Ostanin}}, \bibinfo {author} {\bibfnamefont
  {A.}~\bibnamefont {Ernst}}, \bibinfo {author} {\bibfnamefont
  {I.}~\bibnamefont {Mertig}}, \ and\ \bibinfo {author} {\bibfnamefont
  {W.}~\bibnamefont {Wulfhekel}},\ }\href {\doibase 10.1109/TMAG.2011.2107506}
  {\bibfield  {journal} {\bibinfo  {journal} {Magnetics, IEEE Transactions on}\
  }\textbf {\bibinfo {volume} {47}},\ \bibinfo {pages} {1619} (\bibinfo {year}
  {2011})}\BibitemShut {NoStop}%
\bibitem [{\citenamefont {An}\ \emph {et~al.}(2009)\citenamefont {An},
  \citenamefont {Zhang}, \citenamefont {Fukuyama},\ and\ \citenamefont
  {Yokogawa}}]{An2009}%
  \BibitemOpen
  \bibfield  {author} {\bibinfo {author} {\bibfnamefont {B.}~\bibnamefont
  {An}}, \bibinfo {author} {\bibfnamefont {L.}~\bibnamefont {Zhang}}, \bibinfo
  {author} {\bibfnamefont {S.}~\bibnamefont {Fukuyama}}, \ and\ \bibinfo
  {author} {\bibfnamefont {K.}~\bibnamefont {Yokogawa}},\ }\href {\doibase
  10.1103/PhysRevB.79.085406} {\bibfield  {journal} {\bibinfo  {journal}
  {Physical Review B}\ }\textbf {\bibinfo {volume} {79}},\ \bibinfo {pages} {1}
  (\bibinfo {year} {2009})}\BibitemShut {NoStop}%
\end{thebibliography}%
\end{document}